\documentclass[sigconf, authorversion]{acmart}  %

\usepackage{subfig}
\usepackage{setspace}

\AtBeginDocument{%
  }

\copyrightyear{2024}
\acmYear{2024}
\setcopyright{rightsretained}
\acmConference[ICPE '24 Companion]{Companion of the 15th ACM/SPEC International Conference on Performance Engineering}{May 7--11, 2024}{London, United Kingdom}
\acmBooktitle{Companion of the 15th ACM/SPEC International Conference on Performance Engineering (ICPE '24 Companion), May 7--11, 2024, London, United Kingdom}
\acmDOI{10.1145/3629527.3652276}
\acmISBN{979-8-4007-0445-1/24/05}

\begin{document}
\sloppy

\title[Privacy-Preserving Collaborative Performance Modeling]{Privacy-Preserving Sharing of Data Analytics Runtime Metrics\\for Performance Modeling}%

\iftrue
\author{Jonathan Will}
\email{will@tu-berlin.de}
\affiliation{%
    \institution{Technische Universität Berlin}
    \city{Berlin}
    \country{Germany}
}

\author{Dominik Scheinert}
\email{dominik.scheinert@tu-berlin.de}
\affiliation{%
    \institution{Technische Universität Berlin}
    \city{Berlin}
    \country{Germany}
}

\author{Seraphin Zunzer}
\email{zunzer@campus.tu-berlin.de}
\affiliation{%
    \institution{Technische Universität Berlin}
    \city{Berlin}
    \country{Germany}
}

\author{Jan Bode}
\email{jan.bode@campus.tu-berlin.de}
\affiliation{%
    \institution{Technische Universität Berlin}
    \city{Berlin}
    \country{Germany}
}

\author{Cedric Kring}
\email{c.kring@campus.tu-berlin.de}
\affiliation{%
    \institution{Technische Universität Berlin}
    \city{Berlin}
    \country{Germany}
}

\author{Lauritz Thamsen}
\email{lauritz.thamsen@glasgow.ac.uk}
\affiliation{%
    \institution{University of Glasgow}
    \city{Glasgow}
    \country{United Kingdom}
    \vspace{3mm}
}

\fi

\renewcommand{\shortauthors}{Will et al.}  %

\begin{abstract}
\setstretch{1.0}  %
Performance modeling for large-scale data analytics workloads can improve the efficiency of cluster resource allocations and job scheduling.
However, the performance of these workloads is influenced by numerous factors, such as job inputs and the assigned cluster resources.
As a result, performance models require significant amounts of training data.
This data can be obtained by exchanging runtime metrics between collaborating organizations.
Yet, not all organizations may be inclined to publicly disclose such metadata.

We present a privacy-preserving approach for sharing runtime metrics based on differential privacy and data synthesis.
Our evaluation on performance data from 736 Spark job executions indicates that fully anonymized training data largely maintains performance prediction accuracy, particularly when there is minimal original data available.
With 30 or fewer available original data samples, the use of synthetic training data resulted only in a one percent reduction in performance model accuracy on average.

\end{abstract}

\iftrue
\fi



\maketitle

\iftrue
\pagebreak
\section{Introduction}

Distributed dataflow systems, such as Apache Spark~\cite{zaharia2010spark} and Apache Flink~\cite{carbone2015apache} enable parallel data processing on large clusters of commodity hardware by facilitating parallelization and error handling.
Here, performance models, which accurately estimate a job's runtime on various cluster setups, enable efficient job scheduling and job-specific resource allocation~\cite{islam2021performance, venkataraman2016ernest}.

However, accurately modeling the performance of such data processing jobs is challenging.
Excluding unpredictable events like hardware failures, various factors in the wider execution context influence runtime behavior.
These factors include data analytics algorithm and job parameters, software versions, dataflow framework parameters, input dataset characteristics, the cluster resources provided, and possibly the interference of co-located jobs running on the same cluster~\cite{hsu2018arrow, will2021c3o}.
Many of these factors may vary between job executions.
Therefore, due to the potentially high-dimensional feature space, creating comprehensive performance models necessitates access to substantial amounts of training data.

There are approaches for sharing execution-context-aware performance metrics among collaborators~\cite{will2021c3o, scheinert2023karasu}.
Nevertheless, this approach to collaborative machine learning raises concerns about data privacy, particularly for private sector companies who may be reluctant to share such metadata with competitors.
This applies especially to certain characteristics of their processed datasets that may disclose internal business information, such as a business's customer count.\\
Several methods have been proposed for achieving privacy in collaborative machine learning.
These methods vary in effectiveness depending on the specific application~\cite{li2020federated, goldreich1998secure, fang2021privacy, senavirathne2020role, rankin2020reliability, ping2017datasynthesizer}.
One promising method for sharing training data while maintaining privacy is differential privacy with data synthesis~\cite{rankin2020reliability, ping2017datasynthesizer}.

In this paper, we introduce an automated method for privacy-preserving collaborative performance modeling for dataflow workloads.
Our approach involves obfuscating performance model training data using differential privacy through data synthesis.
Additionally, we assess and discuss the potential of this method to maintain performance model accuracy when using synthetic training data.
We also measure the overhead associated with generating synthetic performance data.
For the evaluation, we use a dataset of 736 unique Spark job executions.

\section{Related Work}

This section explains performance modeling of distributed dataflow workloads and lays out privacy-preserving approaches to collaboratively training machine learning models.

\subsection{Dataflow Job Performance Modeling}

The performance of distributed dataflow jobs is influenced by numerous factors.
These factors include the type and size of the allocated cluster, software versions, along with certain job parameters and dataset characteristics.

Previous works on cluster resource allocation establish performance models for different cluster configurations to learn dataflow job behavior. These models can be employed for automated scheduling or resource allocation decisions~\cite{hsu2018arrow, will2021c3o, islam2021performance, venkataraman2016ernest, scheinert2023karasu, scheinert2021potential}.
\\\emph{Karasu}~\cite{scheinert2023karasu} utilizes shared performance metrics to accelerate the iterative optimization of given objectives like minimizing runtime or carbon emissions.
\\\emph{C3O}~\cite{will2021c3o} shares models and performance metrics for a specific job in a single repository.
The repository maintainers annotate all the job parameters and dataset characteristics that influence performance.

While data sharing approaches can help solve the cold-start problem of model-based performance optimization, a drawback of these approaches is that participating organizations must be willing to share metadata, including dataset characteristics, despite privacy concerns.

\subsection{Privacy in Collaborative Machine Learning}\label{ssec:privacy}

For achieving training data privacy, Liu et al. have identified the following three general categories of approaches~\cite{liu2021when}:

\paragraph{1. Aggregation}
Aggregation-based approaches for privacy in collaborative machine learning involve participants independently training models on their local data and sharing aggregated model updates, such as gradients or statistics instead of raw data.
One prominent example is Federated Learning~\cite{li2020federated}.

\paragraph{2. Encryption}
Training with encrypted data has been demonstrated to be effective for relatively simple models, like Naive Bayes and decision trees~\cite{bost2014machine}.
Notable examples of encryption methods include Secure Multi-Party Computation~\cite{goldreich1998secure} and Homomorphic Encryption~\cite{fang2021privacy}.

\paragraph{3. Obfuscation}
Various obfuscation techniques can be applied to unencrypted training data, including adding noise or generating new data while maintaining statistical properties necessary for training accurate models.
Notable methods include Data Anonymization~\cite{senavirathne2020role} and Data Synthesis~\cite{rankin2020reliability, ping2017datasynthesizer}.%

\vspace{2.5mm}
\hspace{-4mm}
We use an obfuscation-based approach since the other two categories of approaches have shortcomings that limit their applicability to collaborative performance modeling of data analytics workloads.
\\
Aggregation-based methods necessitate continuous cooperation of several collaborators, which might not be feasible for rarely-used data analytics jobs.\\
Encryption-based approaches can share training data, but the limited model viability of those approaches limit their applicability to performance modeling of data analytics workloads.

\pagebreak

\section{Approach}

This section outlines our obfuscation-based approach for privacy-preserving collaborative performance modeling.

\subsection{Idea Overview}

The main aim of our approach is to share performance model training data for data analytics workloads between collaborating organizations.
In order to incentivize data sharing, it is important to anonymize meta information pertaining to a collaborator's workloads.
At the same time, the data must maintain the statistical properties required for training accurate performance models, such as an accurate relation between execution context and runtime.
We are presenting an automated method for collaborative performance modeling of dataflow workloads while preserving privacy, an overview of which is shown in Figure~\ref{fig:figure_one}.

\begin{figure}[h!]
    \vspace{-1.5mm}
    \includegraphics[width=.80\columnwidth, keepaspectratio]{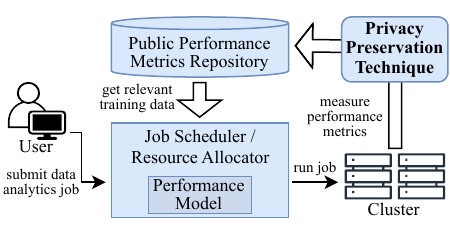}
    \vspace{-4.0mm}
    \caption{High-level overview of privacy-preserving runtime metrics sharing for collaborative performance modeling.}\label{fig:figure_one}
    \vspace{-1mm}
\end{figure}

\vspace{-4mm}
\hspace{.5mm}
\\
\emph{Example Use Case:}\\
An online retailer regularly processes sales data and captures performance metrics, including runtime and execution context, such as the number of rows and columns processed and the public cloud resources used.
From this data, the retailer generates synthetic datapoints that collaborators can use to train reasonably accurate performance models.
Yet, collaborators cannot derive sensitive information from the shared data, such as the number of sales processed by a specific job or the total amount of sales processed during a certain time period.

\subsection{Data Obfuscation via Data Synthesis}

To facilitate privacy-preserving sharing of performance model training data, we employ an obfuscation technique introduced as DataSynthesizer by Ping et al.~\cite{ping2017datasynthesizer}.
This technique generates synthetic data from the original dataset in two steps.
Each step is represented by a separate system module.
\vspace{+2mm}
\\
1) The DataDescriber captures the data types, correlations, and distributions of the attributes in the original dataset and generates a data summary.%
\vspace{2mm}
\\
2) The DataGenerator samples any specified number of synthetic data points from the data distribution summary generated by DataDescriber.
Therefore, the synthetic data points differ from the original data in terms of both quantity and content.
\vspace{2mm}
\\
DataSynthesizer has been released as an open-source tool\footnote{\href{https://github.com/DataResponsibly/DataSynthesizer}{\texttt{github.com/DataResponsibly/DataSynthesizer}}, accessed in January 2024}.

\vfill
\pagebreak

\section{Evaluation}

In this section, we assess the feasibility of our approach through experimental evaluation.
We measure the accuracy of performance models trained with synthetic data and the overhead involved in producing such data.

\subsection{Experimental Setup}\label{ssec:experimental_setup}

In the evaluation, we use the trace dataset and the performance models published in C3O~\cite{will2021c3o}.
\vspace{2mm}
\\
\emph{Trace Dataset.}\\
The dataset contains Spark job executions for five distinct algorithms that were tested across various cluster configurations in Amazon EMR, a managed Spark service.
The algorithms, namely Sort, Grep, Linear Regression, K-Means, and Page Rank, were executed on clusters of different sizes, with varying runtime-influencing job parameters and input dataset characteristics.
The dataset comprises 36, 150, 140, 140, and 270 unique runtime experiments for the aforementioned jobs, respectively.
\vspace{2mm}
\\
\emph{Performance Models.}\\
We utilize the C3O performance modeling system's two default models: Gradient boosting and a model based on Ernest~\cite{venkataraman2016ernest}.
Typically, gradient boosting displays higher accuracy, except in situations with limited availability of training data, where the Ernest model may have superior accuracy.\\
We evaluate the trained model's accuracy with the mean absolute percentage error (MAPE) metric.
For example, if the predicted runtime deviates by 20\%, the MAPE will be expressed as 0.2.
\vspace{2mm}
\\
\emph{Local Hardware.}\\
We measured the overhead of generating synthetic data with a Python script that ran single-threaded on an octa-core AMD Ryzen~7~PRO~4750U processor (1.7-4.1 GHz, 8MB cache).
\vspace{3mm}
\\
The full evaluation is available in a public code repository: \texttt{\href{https://github.com/dos-group/pm-data-privacy}{github.com/dos-group/pm-data-privacy}}

\begin{figure}[b!]
    \centering
    \subfloat{ \includegraphics[width=.47\linewidth]{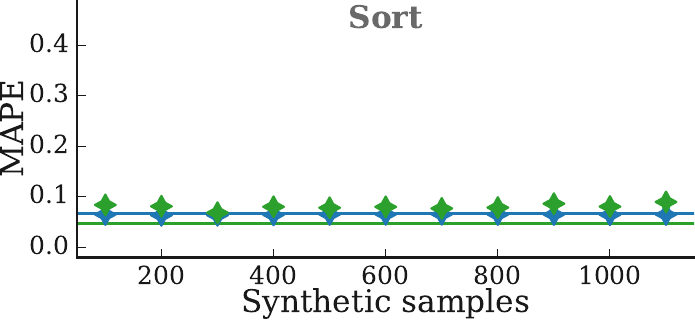} } \hspace{1mm}
    \subfloat{ \includegraphics[width=.47\linewidth]{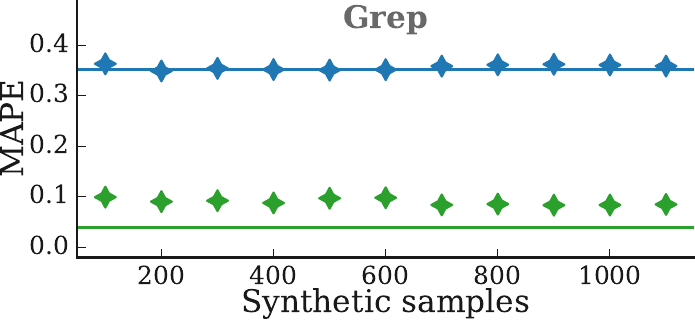} }
    \\\vspace{-2.5mm}
    \subfloat{ \includegraphics[width=.47\linewidth]{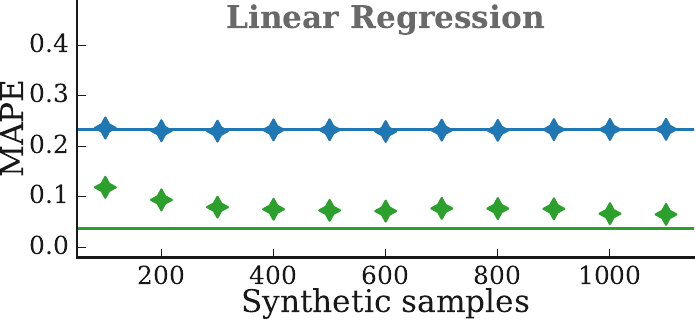} } \hspace{1mm}
    \subfloat{ \includegraphics[width=.47\linewidth]{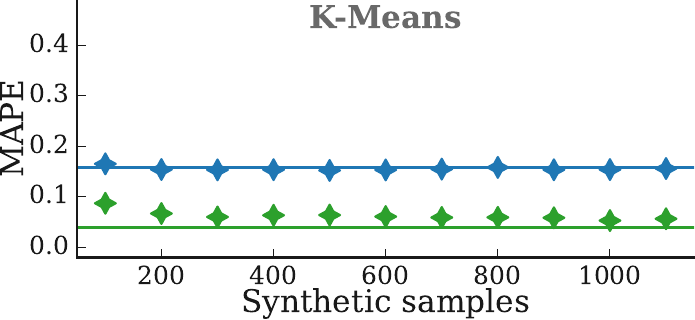} }
    \\\vspace{-2.5mm}\hspace{-3.5mm}
    \subfloat{ \includegraphics[width=.47\linewidth]{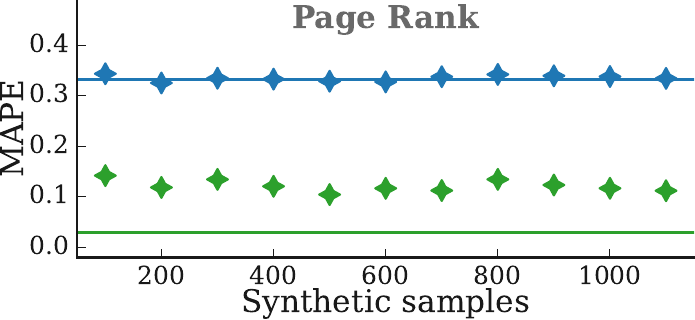} } \hspace{7mm}
    \subfloat{ \includegraphics[width=.37\linewidth]{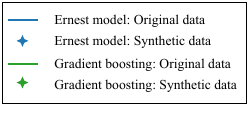} }
    \vspace{-1.0mm}
    \caption{Synthetic training dataset size and resulting performance model error compared to using the full original data.}\label{fig:experiment1}
\end{figure}

\vfill
\pagebreak

\begin{figure}[b!]
    \centering
    \subfloat{ \includegraphics[width=.47\linewidth]{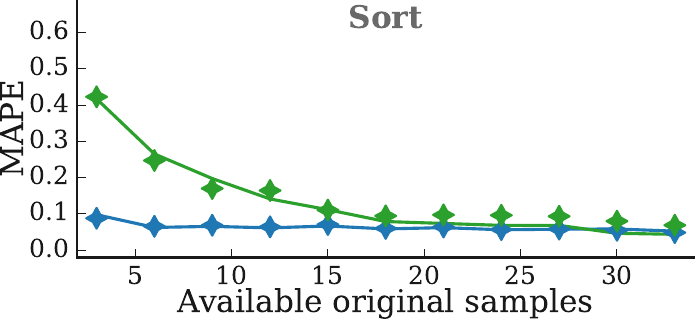} } \hspace{1mm}
    \subfloat{ \includegraphics[width=.47\linewidth]{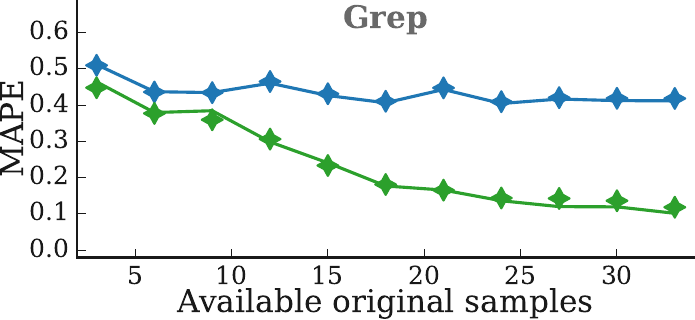} }
    \\\vspace{-2.5mm}
    \subfloat{ \includegraphics[width=.47\linewidth]{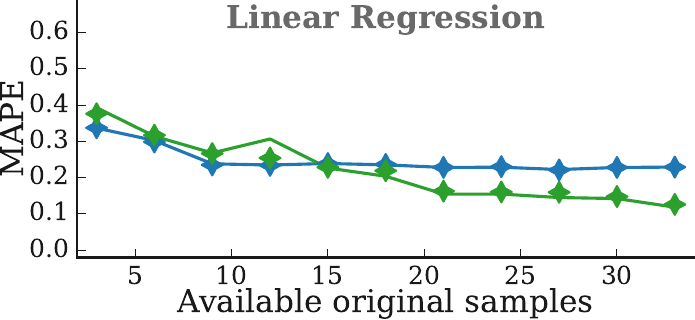} } \hspace{1mm}
    \subfloat{ \includegraphics[width=.47\linewidth]{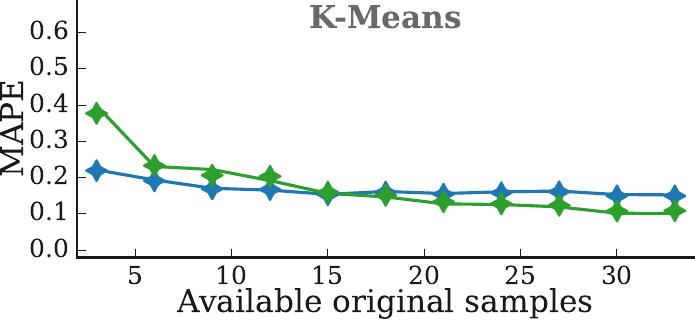} }
    \\\vspace{-2.5mm}\hspace{-3.5mm}
    \subfloat{ \includegraphics[width=.47\linewidth]{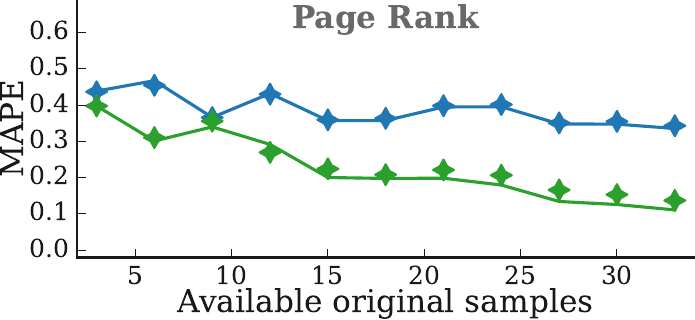} } \hspace{7mm}
    \subfloat{ \includegraphics[width=.37\linewidth]{images/legend.pdf} }
    \vspace{-1.0mm}
    \caption{Performance model error with 1000 synthetic data samples, generated from small amounts of original data.}\label{fig:experiment2}
\end{figure}

\subsection{Performance Modeling with Synthetic Data}

A viable approach must allow for accurate performance modeling with synthetic data.
We assess the performance model's error trained on original and synthetic data in various scenarios.
\vspace{2mm}
\\
\emph{Synthetic Training Data Size and Performance Model Accuracy.}
\vspace{1mm}
\\
First, we investigate how the creation of a substantial quantity of synthetic training data impacts model accuracy.
We measured the accuracy of performance models that were trained on complete original datasets for different Spark jobs. %
Then, we extracted various amounts of synthetic data from the same dataset, mainly larger quantities than the original data available.
With this sample data, we retrained the models and measured their accuracy.

Figure~\ref{fig:experiment1} shows the results of this experiment.
They suggest that additional synthetic training data does not have an observable impact on model accuracy beyond a certain point.
Further, using synthetic data for training works differently well compared to using original data, depending on the model performance and type of job.
\vspace{-2mm}
\\
\emph{Sampling Synthetic Data from Few Available Original Data Points.}
\vspace{1mm}
\\
Next, we examine the feasibility of generating synthetic training data with limited availability of original training data.
To this end, we randomly selected different small quantities of samples from the original performance dataset and generated 1000 synthetic data points from that.
Then, we trained the performance models on both the original and synthetic data and compared their prediction accuracy.

Figure~\ref{fig:experiment2} shows the results of this experiment.
We find that when original data availability is \emph{low}, the use of synthetic data for training results in performance models that are nearly as accurate.
For availability ranging from 3 to 30 original samples, we observed a difference of only 1.14\%.
However, this contrasts with the results from the previous experiment shown in Figure 2, where models trained on synthetic data sampled from a \emph{larger} dataset of original data can perform significantly worse than models trained on the original data.

\vfill
\pagebreak

\subsection{Data Synthesis Overhead}

A viable approach to creating synthetic data for accurate performance modeling must not impose excessive overhead for data creation.
We measured the overhead of creating synthetic training data on the hardware described in Section~\ref{ssec:experimental_setup}.

\begin{figure}[h!]
    \centering
    \subfloat{ \includegraphics[width=.43\linewidth]{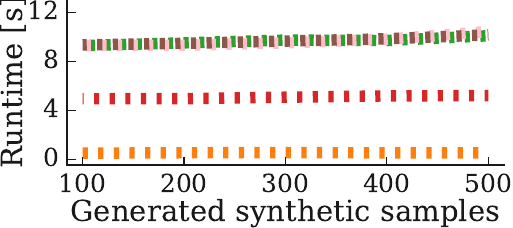} } \hspace{3mm}
    \subfloat{ \includegraphics[width=.43\linewidth]{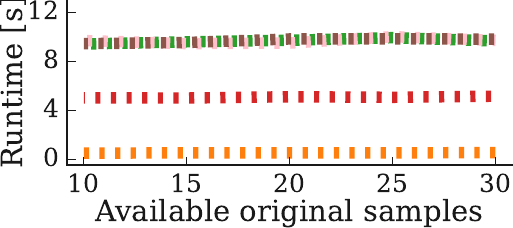} }
    \\\vspace{-1.0mm}
    \subfloat{ \includegraphics[width=1\linewidth]{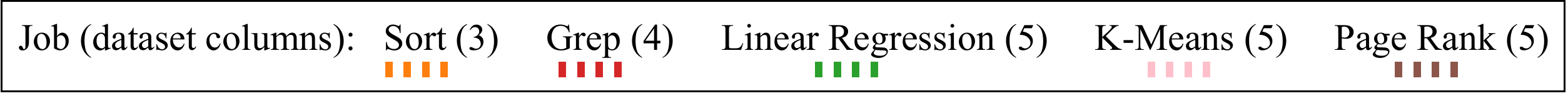} }
    \vspace{-0.5mm}
    \caption{Overhead for creating synthetic data for different Spark job performance datasets.}\label{fig:experiment3}
\end{figure}

In Figure~\ref{fig:experiment3}, we see that for performance datasets containing the runtime and runtime-influencing factors of typical Spark jobs, this overhead was measured to be approximately between half a second and ten seconds.
We observe that the computational cost of synthesizing data does not significantly increase with an increase in the amount of sampled synthetic data or the number of available samples in the original dataset.
Rather, the findings suggest that the primary computational effort arises from processing each attribute, i.e., column, in the original dataset.
In the case of DataSynthesizer, this part is conducted by the DataDescriber component.

\vspace{2mm}
\subsection{Discussion}

We will now discuss the experimental evaluation's results in terms of the practical implications for our approach's viability.

First, we observed that it is feasible to generate substantial quantities of synthetic data without compromising the model's accuracy.
This implies that we can achieve privacy not just by modifying the content of each data point, but also by creating arbitrarily large amounts of synthetic data, thereby concealing the actual quantity of processed jobs.

Then, it has been observed that the model accuracy gap when using synthetic data is lowest when the quantity of original data points is low.
In instances where publicly shared training data points are unavailable or rare, the introduction of synthetic data can have a significant positive effect on the model accuracy of collaborators.
Consequently, sharing synthetic data is particularly advantageous in the early stages of a training data sharing initiative.

Finally, the computational overhead of generating synthetic performance data has been shown to range in seconds for performance datasets of typical Spark jobs on typical consumer hardware.
This low amount of time should not discourage collaborators from generating and sharing synthetic data.

\vfill
\pagebreak

\section{Conclusion}\label{sec:CONCLUSION}

In summary, this paper has explored how differential privacy via data synthesis can facilitate the sharing of runtime data for performance modeling of data analytics workloads in a privacy-preserving manner.
Our initial method has demonstrated an acceptable trade-off between model prediction accuracy and data privacy.
Especially in cases where there is limited available performance data overall, the accuracy of collaborators' performance models can be significantly improved through the use of shared synthetic training data samples.
Further, the data synthesis has been shown to induce low computational overhead.

In the future, we will investigate alternative approaches to ensure privacy when sharing performance metrics of data analytics workloads.
Moreover, we hope our short paper also inspires further research by others in the same direction.

\fi

\begin{acks}
\iftrue
This work has been supported through a grant by the German Research Foundation (DFG) as ``C5'' (grant 506529034).
\fi
\end{acks}

\bibliographystyle{ACM-Reference-Format}

\bibliography{./references}  %

\end{document}